\documentclass[12pt]{article}
\usepackage{epsfig}

\begin{document}
\begin{titlepage}
\begin{center}

{\Large\bf Transport approach to the superconducting proximity
effect in carbon nanotubes}
\end{center}
\vskip 0.6truecm
\begin{center}
{\large\bf J. Gonz\'alez\footnote{imtjg64@pinar2.csic.es }}
  \vspace{0.5cm} \\
{\it Instituto de Estructura de la Materia,\\
         Consejo Superior de Investigaciones Cient{\'\i}ficas,\\
Serrano 123, 28006 Madrid, Spain} \vspace{0.5cm}
\end{center}

\begin{abstract}
A microscopic approach is developed to account for
the magnitudes of the supercurrents observed experimentally
in carbon nanotubes placed between superconducting contacts.
We build up a model for the nanotube ropes encompassing the
electron repulsion from the Coulomb interaction and the
effective attraction given by phonon exchange. We show that
the available experimental data are consistent with the
expected decay of the supercurrents along the length of the
nanotube samples. Our results stress that the propagation
of the Cooper pairs is favored in the thick ropes, as a
consequence of the reduction in the strength of the Coulomb
interaction from the electrostatic coupling between the
metallic nanotubes. We also provide an explanation for the
temperature dependence of the supercurrents observed in the
experiments, remarking the existence of a crossover from a
very flat behavior at low temperatures to a pronounced
decay in the long nanotubes.

\end{abstract}

\vspace{4cm}
\begin{flushright}
{\em Submitted to J. Phys.: Condens. Matter}\\
{\em (special issue on Nanostructured Surfaces)}
\end{flushright}

\end{titlepage}

\section{Introduction}

Since their discovery in 1991, the carbon nanotubes have shown
remarkable electronic properties. They may display quite different
behaviors, depending on their geometry and the quality of the
contacts used to probe the electronic transport. It was first
predicted\cite{hso,sdd,mint}, 
and later checked experimentally\cite{exp1,exp2}, 
that the nanotubes
have semiconducting or metallic behavior depending on the
helicity of the hexagonal carbon rings wrapped around the
tubule. This is the kind of versatility that points at the
carbon nanotubes as ideal molecules to build electronic devices
at the nanometer scale\cite{ency}.

The electron correlations are moreover very strong in the carbon
nanotubes, what leads to the breakdown of the conventional
picture of electronic transport at low energies. 
The Coulomb interaction prevails in the
individual nanotubes\cite{fisher,eg}, and it
drives the electron system to a state with the properties of the
so-called Luttinger liquid, characterized by the absence of
quasiparticles at the Fermi level.  This feature reflects in the
power-law behavior of observables like the tunneling density of
states, whose suppression at low energies has been actually
observed in the measurements of the conductance in the carbon
nanotubes\cite{mac,yao}.

There have been also experiments revealing the existence of 
superconducting correlations in the nanotubes\cite{kas2,marc}. The
first observations were made in nanotube ropes suspended between
superconducting electrodes, the most remarkable signal being the
appearance of supercurrents for temperatures below the critical
value of the contacts\cite{kas2}. What has been measured in that 
experiment is the proximity effect, by which Cooper pairs
are formed in the nanotubes near the superconducting contacts.
Later, superconducting transitions have been observed in ropes
suspended between metallic, nonsuperconducting electrodes\cite{kas}. 
The measurements reported in Ref. \cite{kas} point at the existence 
of a superconducting phase intrinsic to the carbon nanotubes. More
recently, strong superconducting correlations have been also
reported in individual nanotubes of very short radius, inserted
in a zeolite matrix\cite{chin}.

The observation of superconducting correlations in the nanotubes
implies that there is a regime which is different to that marked
by the dominance of the Coulomb interaction. The appearance of
the superconducting effects requires a suitable attenuation of
the repulsive interaction within the nanotubes, due to either
the electrostatic coupling among the metallic nanotubes in a
rope or the coupling with nearby conductors. Another factor that
seems to be crucial is the achievement of high quality contacts
between the nanotubes and the electrodes. This has been possible
in the experiments described in Ref. \cite{kas2} by the use of a
technique that allows to sold and suspend the nanotubes between
the contacts. The electrodes were made of a Re/Au bilayer (with
transition temperature $T_c \approx 1.1 \; {\rm K}$) in the case
of a massive rope, and of a Ta/Au bilayer (with $T_c \approx 0.4
\; {\rm K}$) in the case of a thin rope.  The room-temperature
resistance of the ropes was consistent in each case with a
resistance of the constituent metallic nanotubes below $h/e^2
\approx 25.8 \; {\rm k}\Omega $ \cite{kas2}, which is of the 
order of the value $h/4e^2$ corresponding to ballistic transport 
in a nanotube.

Below the transition temperature of the contacts, the nanotubes
can support currents without developing any resistance\cite{kas2}. 
These
so-called supercurrents reach a maximum value $I_c$, which is
known as the critical current. This has been measured for the
samples described in Ref. \cite{kas2}  and, as pointed out there, 
the magnitudes and temperature dependences obtained cannot be
understood in the framework of the conventional proximity
effect. The critical current is usually related to the values of
the normal resistance $R_N$ and the energy gap $\Delta $ in the
superconductor through the expression $I_c = (\pi /2) \Delta /e
R_N$. However, the value estimated in this way turns out to be
about 40 times smaller than the experimental measure for the
thin rope reported in Ref. \cite{kas2}. The observed dependence 
of the critical current on temperature is also very flat in that
sample, before getting close to the transition temperature of
the electrodes. This is again at odds with the conventional
picture, in which the critical current should reflect the
temperature dependence of the superconducting gap at the contacts.

A consistent explanation of the behavior of the supercurrents
measured experimentally has been proposed in Ref. \cite{jose1}. 
It has been stressed there the need to consider the supercurrents 
as an effect of one-dimensional (1D) transport along
the carbon nanotubes. The propagation of the Cooper pairs in the
nanotubes is favored in the thick ropes, as there is
a larger attenuation in the strength of the Coulomb interaction
for higher content of metallic nanotubes in a rope\cite{jose2}. 
The origin of this effect is similar to that of screening in a
three-dimensional (3D) conductor. In a rope, however, the tunneling
amplitude between neighboring metallic nanotubes is highly 
suppressed in general, due to the misalignment of the respective 
carbon lattices\cite{mkm,avour}. 
The absence of a significant intertube electron hopping is what
keeps the Coulomb interaction long-ranged in the nanotubes,
despite the large reduction of its strength in the thick ropes.

In the present paper, we confront the results
from a microscopic model of the nanotube ropes with the
experimental measures from Ref. \cite{kas2}. The propagation 
of the Cooper pairs along the carbon nanotubes gives rise to
supercurrents with a strong dependence on the length of the
junction, when the Coulomb interaction prevails in the
electron system. Thus, the measures of the supercurrents, 
available in two experimental samples with different
length and content of metallic nanotubes, can be used to discern
the character of the interactions present in the nanotubes. 
We will see, in particular, that the value of the critical
current in the thin rope reported in Ref. \cite{kas2} can be 
accounted for by considering an attractive component in the
electron-electron interaction, coming from the coupling of
the electrons to the elastic modes of the carbon lattice. 
In general, the balance between the renormalized Coulomb
interaction in the ropes and the attractive component of the
interaction allows to explain the magnitudes as well as the
temperature dependence of the critical currents reported in
Ref. \cite{kas2}.

In the next section, we review the different sources of 
electron-electron interaction in the carbon nanotubes and 
describe the microscopic model relevant for the nanotube ropes.
Section 3 is devoted to set up the framework used for the 
study of transport along the junctions. The comparison between
the results from the microscopic model and the experimental 
data is carried out in Section 4. Finally, the conclusions of
our investigation are drawn in Section 5.

\section{Electronic interactions in carbon nanotubes}

\begin{figure}
\begin{center}
\raisebox{1cm}{\epsfxsize 6cm \epsfbox{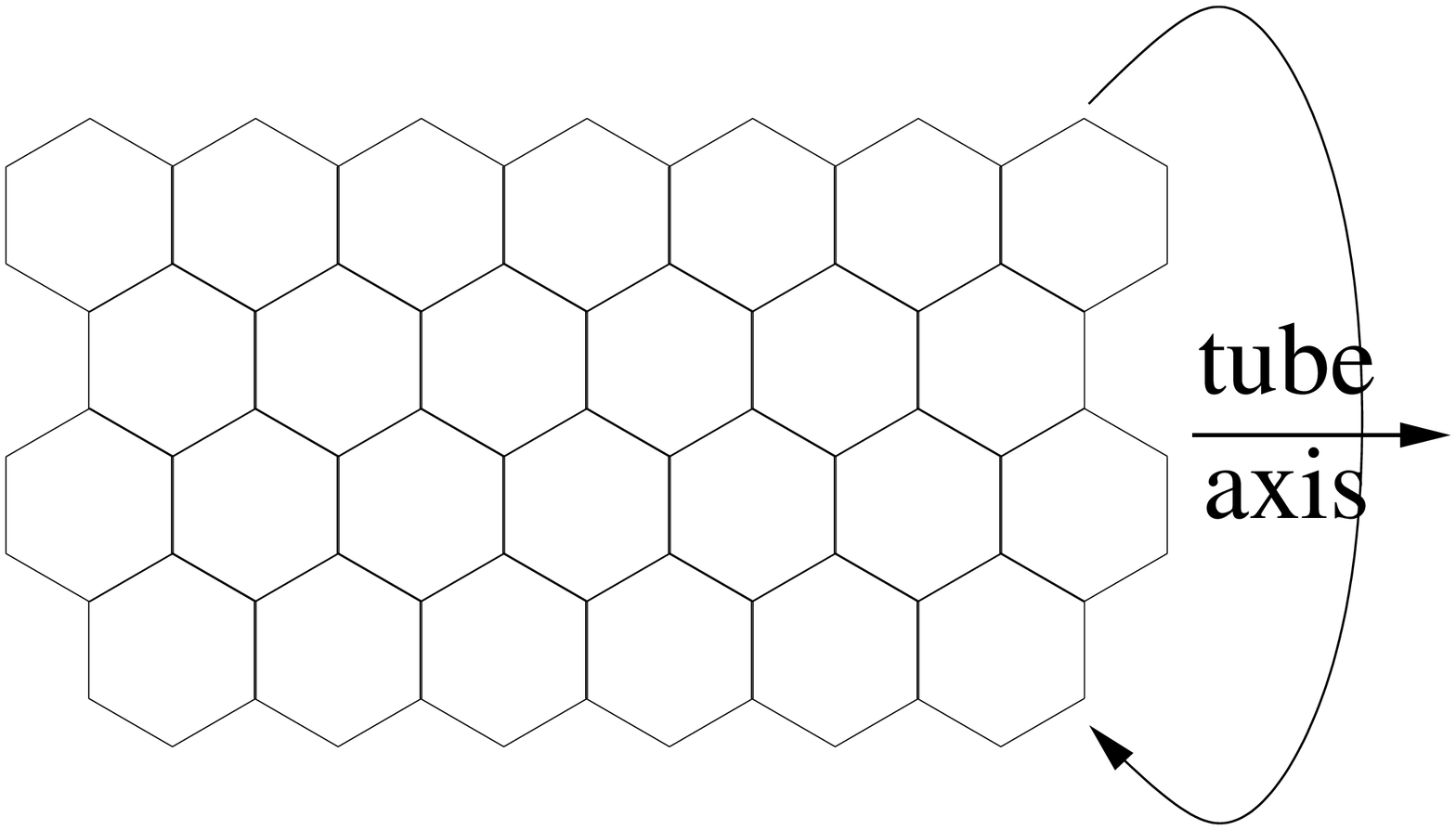}}
\hspace{1cm} \mbox{\epsfysize 6cm \epsfbox{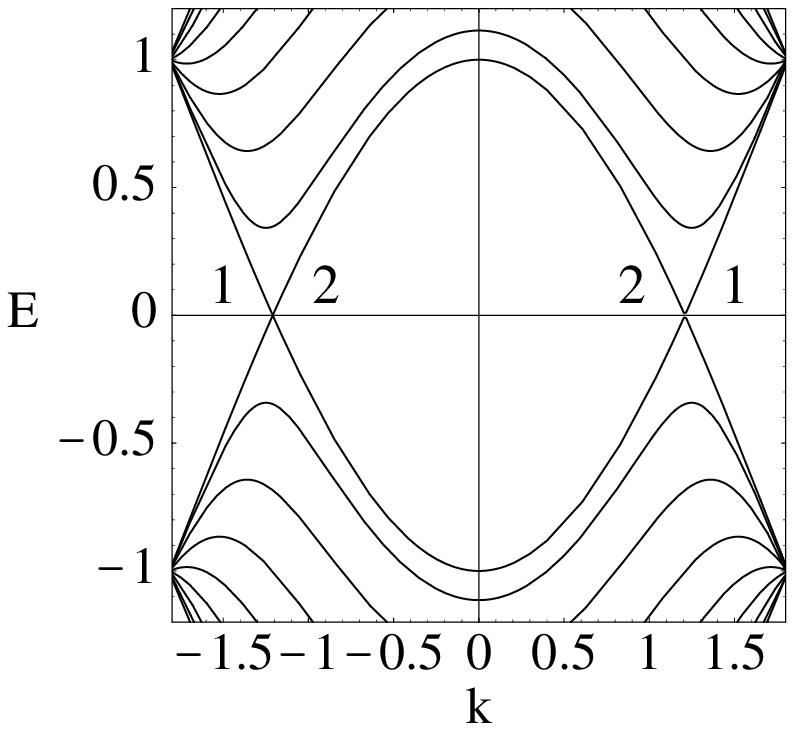}}
\end{center}

\caption{Left: Schematic representation of the wrapping action
leading to an armchair nanotube. Right: Band structure of an armchair 
nanotube with nine subbands. The energy is
measured in units of the overlap integral and the momentum in
units of the inverse of the C-C distance.}

\label{one}
\end{figure}

We review first the structure of the modes and interactions that
take place at low energies in the carbon nanotubes. The band
structure of all the metallic nanotubes has two pairs of linear
branches that cross at opposite Fermi points (for undoped
nanotubes) as shown in Fig. \ref{one}. For typical
nanotubes in a rope with radius $R$ around $\approx 0.7$ \AA,
the electron modes in the energy range below $\approx 0.5 \; {\rm
eV}$ belong to the linear branches near the Fermi level, as can
be seen from Fig. \ref{one}.  Therefore, we will pay attention 
only to them when studying the low-energy properties of
the carbon nanotubes.

We will carry out the discussion throughout this section having
in mind the specific geometry of the armchair nanotubes, 
illustrated in Fig. \ref{one}, while a
completely parallel construction can be done for the metallic
chiral nanotubes.  The linear branches found near the Fermi
level can be divided in two classes, according to the different
symmetry of their modes under the exchange of the two
sublattices of the hexagonal carbon lattice.  In the case of
armchair nanotubes, the two subbands that cross at the Fermi
level, labeled 1 and 2 in Fig.
\ref{one}, correspond to the eigenvalues of the one-particle
hamiltonian
\begin{equation}
{\cal H} =   t  \left(
\begin{array}{cc}
0  &   1 - 2 \cos (\sqrt{3} k a/2)    \\ 1 - 2 \cos (\sqrt{3} k
a/2)    &   0
\end{array}         \right)
\end{equation}
that operates in the space of the electron amplitudes in the two
different atoms of the lattice basis\cite{nucl}. The parameter $t$
represents the matrix element of the potential for
nearest-neighbor carbon orbitals and $a$ is the C-C distance.
Then, it is clear that the modes in subband 1 have a smooth
amplitude in the carbon lattice, while the modes in subband 2
have an amplitude that alternates the sign when shifting between
nearest-neighbor atoms. This has important consequences
regarding the form of the interactions between the different
branches.

\subsection{Coulomb interactions}

The Coulomb interaction provides a strong source of repulsion
between the electrons in the carbon nanotubes. One may obtain an
effective 1D potential $V_C (x-x')$ on the longitudinal
dimension of the nanotube, by taking the average of the 3D
interaction over two nanotube sections placed at a spatial
separation $x-x'$ \cite{eg}. The Fourier transform with respect 
to this variable gives a potential which is logarithmically 
singular for small momentum transfer $k$ \cite{wang},
\begin{equation}
V_C (k) = (e^2 /2\pi ) \log |(k_c + k)/ k|
\end{equation}
The large momentum $k_c$, of the order of $\sim 1/R$, survives
in the projection onto the longitudinal dimension as a remnant
of the finite radius of the nanotube.

Actually, the Coulomb interaction remains singular only when
acting between currents that belong each to a given linear
branch.  Otherwise, if one of the electron modes is scattered to
another branch with different symmetry in the two-atom basis,
the product of the {\em in} and {\em out} amplitudes has an
alternating sign over the hexagonal carbon lattice. Then, the
scattering amplitude averages to zero when making the 1D
projection with a large spatial separation $x-x'$ along the
nanotube.

At short distances, the Coulomb interaction may still give rise
to a residual 1D interaction mixing {\em in} and {\em out} modes
in different linear branches, but these processes do not bear
the singular character of the long-range interaction. This also
applies to the processes with large momentum transfer in which
the electron modes are scattered from one of the Fermi points to
the other. These backscattering and Umklapp processes
have a strength which is reduced by a factor of the order of
$\sim 0.1 a/R$ with respect to the nominal strength $e^2/2 \pi $
of the long-range Coulomb interaction\cite{fisher,eg}. 
For nanotubes with $R
\approx  0.7$ \AA, that factor has a magnitude of $\sim 0.01$.
Thus, the backscattering and Umklapp interactions of this kind
turn out to be negligible in comparison to the Coulomb interaction
between currents in definite linear branches.

\subsection{Electron-phonon interactions}

Another source of electron-electron interaction comes from the
coupling of the electrons to the elastic modes of the carbon
lattice. The exchange of phonons gives rise to a retarded
interaction, which can be represented in terms of an effective
potential $V(k,\omega)$ of the form
\begin{equation}
V (k,\omega ) = - g_{p,p'}(k) g_{q,q'}(-k) \frac{\omega_k }{-
\omega^2 + \omega_k^2 }
\label{pot}
\end{equation}
where $\omega_k $ is the phonon energy and the $g_{p,p'}(k)$ are
appropriate electron-phonon couplings. Thus, for frequencies
$\omega $ below the characteristic phonon energies, the
potential $V(k,\omega)$ may reach negative values, providing
then a kind of effective attraction between the electrons.

The strength of the effective interaction depends on the
electron-phonon couplings $g_{p,p'}(k)$, where the indices $p$
and $p'$ label the gapless subbands to which the {\em in} and
{\em out} electron modes belong.  For the computation of the
$g_{p,p'}(k)$, we rely on the tight-binding approximation, which
is fairly appropriate for the carbon nanotubes. The
electron-phonon couplings can be represented by a sum over
nearest neighbors of the atoms in the unit cell of the 
nanotube\cite{jdd}
\begin{equation}
g_{p,p'} (k-k')   =   \frac{1}{( \mu \; \omega_{k-k'} )^{1/2}
} \sum_{\langle s , s' \rangle } u_s^{(p) *} (k)   u_{s'}^{(p')}
(k')        ( \mbox{\boldmath $\epsilon$}_s
(k-k') - \mbox{\boldmath $\epsilon$}_{s'} (k-k') )
\mbox{\boldmath $\cdot \nabla$}   J (s,s')
\label{tight}
\end{equation}
where $\mbox{\boldmath $\epsilon$}_s (k-k')$ is the phonon
polarization vector at site $s$, $u_{s'}^{(p')} (k')$ and
$u_s^{(p)} (k)$ are the respective amplitudes of the incoming
and outgoing electrons, $J(s,s')$ is the matix element of the
potential between orbitals at $s$ and $s'$, and $\mu $ is the
mass per unit length. We are already implying that the coupling
can be expressed as a function of $k-k'$, what is a reasonable
approximation for small values of that momentum transfer (modulo
$2 k_F$).

The product of the two couplings in (\ref{pot}) can be actually
positive or negative, depending on the linear branches 
involved\cite{jdd}.
The sign is dictated by symmetry rules which are direct
consequence of the geometry of the nanotube lattice. We
illustrate that fact for the case of the armchair nanotubes,
though a similar argument can be developed in general for
metallic chiral nanotubes. Thus, taking into account that the
modes with antibonding character have an amplitude with
alternating sign in nearest-neighbor sites, we observe from
inspection of (\ref{tight}) that
\begin{equation}
g_{1,1} (k) = - g_{2,2} (k)
\label{long1}
\end{equation}
The electrons can be also scattered from one of the subbands to
the other by exchanging phonons. The fact that one of the modes
has then antibonding character reflects in the antisymmetric
condition
\begin{equation}
g_{1,2} (k) = - g_{2,1} (k)
\label{long2}
\end{equation}
The relations (\ref{long1}) and (\ref{long2}) are based on
symmetry rules, and they hold irrespective of whether the
exchanged phonons are optical or acoustic.

Using the fact that $g_{p,p'}^{*} (k-k') =  g_{p',p} (k'-k)$, it
is clear from (\ref{pot}) that the exchange of phonons gives
rise to an attractive electron-electron interaction when the
electrons are scattered within the same subband.  On the other
hand, the condition (\ref{long1}) on the couplings makes the
interaction repulsive between electron currents in different
subbands. These considerations hold for frequencies below the
typical Debye energy of the phonons, which is in the range
between 0.1 and 0.2 eV for nanotubes with radius around $\approx
0.7$ \AA \cite{span,mahan,saito}.  
Taking into account these magnitudes, one can make an
estimate of the strength of the effective interaction, which
turns out to be of the order of $\sim (\partial t/\partial
a)^2/\mu \omega_k^2
\sim 0.1 v_F$.

\subsection{Low-energy dynamics}

When describing the low-energy dynamics of the electrons, one
has to bear in mind the special structure of the nanotube ropes.
These are made of a disordered mixture of nanotubes
with different helicities and diameters. In these conditions,
the tunneling of electrons between neighboring metallic
nanotubes is highly suppressed. This comes from the fact that
the carbon lattices of neighboring nanotubes are not aligned in
general, what leads to a mismatch between the respective Fermi
points and to the consequent difficulty to conserve the
longitudinal momentum in the tunneling process\cite{mkm}. 
An experimental
signature of the very small tunneling amplitude has been
observed in the measurement of the coupling resistance between
tubes, which has shown wide variations ranging between $2 \;
{\rm M}\Omega $ and $140 \; {\rm M}\Omega $ in different
samples\cite{avour}. 
This is consistent with a picture in which the tunneling 
amplitude between nanotubes may be about three orders of magnitude 
below the electron hopping amplitude within the carbon lattice.
We will therefore assume that the coupling between nanotubes is
given by the Coulomb interaction, and that the kinetic energy of
the electrons can be represented by the sum of the kinetic
energies of the metallic nanotubes in the rope.

\begin{figure}
\begin{center}
\mbox{\epsfxsize 7cm \epsfbox{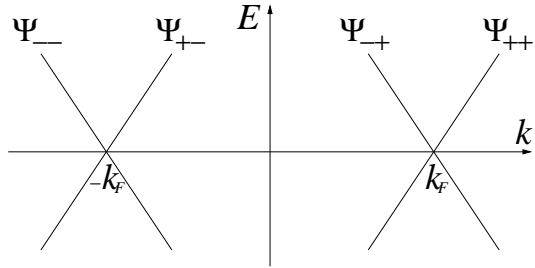}}
\end{center}
\caption{Linear branches and respective electron fields in the 
low-energy spectrum of a metallic nanotube.}
\label{two}
\end{figure}

Within each nanotube, one has to balance the repulsive Coulomb
interaction and the effective electron-electron interaction
coming from phonon exchange. As remarked above, there are no
significant backscattering or Umklapp processes coming from the
Coulomb interaction, so that the competition between that and
the effective attraction takes place in processes where the
scattered electrons remain in their respective branches. Our
approach will focus on the description of this balance between
repulsive and attractive interactions.
For that purpose, we will build a model in which
the elementary objects are the electron density operators, given
in terms of the Fermi fields $\Psi^{(a)}_{r i \sigma }(x)$ for
the different linear branches (shown in Fig. \ref{two}) by
\begin{equation}
 \Psi^{(a) \dagger}_{r i \sigma}(x) \Psi^{(a)}_{r i \sigma }(x)
= \rho^{(a)}_{r i \sigma }(x)
\end{equation}
The label $a$ runs over the different metallic nanotubes in a
rope, $a = 1, \ldots n$. The index $r = \pm $ is used to label
the left- or right-moving character of the linear branch, and
the index $i = \pm $ to label the Fermi point.  In this way, we
are setting aside the backscattering processes, which tend
anyhow to enhance the superconducting correlations\cite{jose3}. 
Then, it can be
thought that our model leads to a slight underestimation of
these effects, for a given strength of the bare attractive
interaction. This can be otherwise corrected by a suitable
renormalization of the effective coupling derived from phonon
exchange, assuming that it reaches greater values in the
low-energy regime of the model\cite{prl}.

According to the above considerations, we write the hamiltonian
for a nanotube rope in the form
\begin{eqnarray}
H_0  & = &    \frac{1}{2} v_F \int_{-k_c}^{k_c} dk \sum_{a r i
\sigma }  : \rho^{(a)}_{r i \sigma} (k) \rho^{(a)}_{r i \sigma}
(-k)  :   \nonumber      \\ &  &    + \frac{1}{2}
\int_{-k_c}^{k_c} dk \; \sum_{a r i \sigma } \rho^{(a)}_{r i
\sigma} (k) \; \sum_{b s j \sigma'  }  V^{(ab)}_{r i, s j} (k)
\; \rho^{(b)}_{s j \sigma'} (-k)
\label{ham}
\end{eqnarray}
The first term in (\ref{ham}) comes from the known representation
of the kinetic energy for a 1D electron system\cite{emery,sol}.
The potential $V^{(ab)}_{r i, s j} (k)$ mediates the
interactions within a given nanotube for $a = b$ and between
different nanotubes for $a \neq b$. Due to the long-range
character of the Coulomb interaction, its contribution has to be
included in all the elements $V^{(ab)} $. It is only for $a =
b$, however, that one has to add the contributions arising from
phonon exchange within each nanotube.

We can introduce two plausible assumptions that lead to a
straightforward resolution of the model under study. The first
is that all the interactions in the nanotube system are
spin-independent. Then, it is appropriate to introduce the
symmetric and antisymmetric combinations of density operators
for the two spin projections. Only the charge density operators
\begin{equation}
\rho^{(a)}_{r i \rho }(k)  = \frac{1}{\sqrt{2}}
 \left(  \rho^{(a)}_{r i \uparrow }(k) + \rho^{(a)}_{r i
\downarrow }(k)   \right)
\end{equation}
enter in the interaction term of the hamiltonian (\ref{ham}).
This implies, in particular for the calculation of the
correlators, that half of the degrees of freedom in the model
are not affected by the interaction.
 
We will further assume that the strength of the effective
interaction from phonon exchange does not depend on the
particular linear branches to which the scattered electron modes
belong. We know however that, if the interaction takes place
between currents in different subbands, the condition
(\ref{long1}) introduces a relative minus sign. It is convenient
to define the respective charge density operators for the
bonding and the antibonding subband
\begin{eqnarray}
\tilde{\rho}^{(a)}_{1 \rho }(k) & = &
                  \rho^{(a)}_{+ + \rho }(k) + \rho^{(a)}_{- -
\rho }(k)    \label{1}        \\
\tilde{\rho}^{(a)}_{2 \rho }(k) & = &
                   \rho^{(a)}_{+ - \rho }(k) + \rho^{(a)}_{- +
\rho }(k)
\label{2}
\end{eqnarray}

The Coulomb interaction has always repulsive character, and it
acts on the symmetric combination of (\ref{1}) and (\ref{2})
\begin{equation}
\tilde{\rho}^{(a)}_{+ \rho }(k)  =    \frac{1}{\sqrt{2}}
   \left(   \tilde{\rho}^{(a)}_{1 \rho }(k) +
\tilde{\rho}^{(a)}_{2 \rho }(k)   \right)
\end{equation}
The effective interaction from phonon exchange is attractive
between currents in the same subband, but repulsive for currents
in different subbands. It therefore acts on the antisymmetric
combination
\begin{equation}
\tilde{\rho}^{(a)}_{- \rho }(k)  =   \frac{1}{\sqrt{2}}
    \left(    \tilde{\rho}^{(a)}_{1 \rho }(k) - \tilde{\rho
}^{(a)}_{2 \rho }(k)   \right)
\end{equation}

In the new density variables, the hamiltonian can be written in
the form
\begin{eqnarray}
H_0  & = &    \frac{1}{2} v_F \int_{-k_c}^{k_c} dk \sum_{a r i
\sigma }  : \rho^{(a)}_{r i \sigma} (k) \rho^{(a)}_{r i \sigma}
(-k)  :   \nonumber    \\
&  &    + \frac{1}{2}
\int_{-k_c}^{k_c}  \frac{dk}{2\pi } \; \left(  4    \sum_a
\tilde{\rho}^{(a)}_{+ \rho} (k) \; V_C (k)   \sum_b     \;
\tilde{\rho}^{(b)}_{+\rho} (-k)   \right.     \nonumber     \\
&  &  \left. \;\;\;\;\;\;\;\;\;\;\;\;\;\;\;\;\;\; +  4  g  \sum_a
\tilde{\rho}^{(a)}_{- \rho} (k) \; \tilde{\rho}^{(a)}_{- \rho}
(-k)    \right)
\label{ham2}
\end{eqnarray}
where $g \; (< 0)$ parametrizes the strength of the effective
interaction from phonon exchange.  The hamiltonian (\ref{ham2})
encompasses the main interactions in the nanotube system,
providing a suitable starting point to study the superconducting
correlations in the nanotube ropes.

\section{Superconducting correlations and supercurrents in 
carbon nanotubes}

We introduce in what follows the theoretical framework
appropriate for the description of supercurrents in the carbon
nanotubes. The study of the Josephson effect in a Luttinger
liquid placed between macroscopic superconductors has been
carried out in Refs. \cite{prox3} and \cite{prox4}.
We first review briefly the approach proposed in Ref.
\cite{prox4} for the calculation of the critical current through
that kind of junction.

In order to describe the coupling to superconducting electrodes,
one more term $H_{Tj}$ has to be added to the hamiltonian of the
1D system for each of the tunnel junctions created at the
contacts.  The term $H_{Tj}$ expresses the hopping of electrons
from the 1D system to the superconductor and viceversa, with a
given tunneling amplitude $t_j$. It has then the form
\begin{equation}
H_{Tj} =  t_j \sum  \Psi^{(a) \dagger}_{ \sigma }(d_j) \Psi_{Sj,
\sigma}(d_j)  \; + \; {\rm h.c. }
\end{equation}
where $\Psi_{Sj, \sigma}(d_j) $ is the electron field at the end
of the superconductor $Sj$, and $\Psi^{(a)}_{ \sigma }(d_j)$ is
a generic electron field of the 1D system.

On top of that, one has also to include the hamiltonians
describing the condensates for the respective macroscopic
superconductors.  For our purposes, all the relevant information
about their properties can be given in terms of the respective
order parameters $\Delta_j$ and normal densities of states $N_j$
at the contacts.

The supercurrent $I$ through the 1D system can be computed in
terms of the derivative of the free energy ${\cal F}$ with
respect to the phase difference $\chi $ of the order
parameters\cite{prox4}
\begin{equation}
I (\chi ) = - 2e \frac{\partial {\cal F}}{\partial \chi}
\label{cur}
\end{equation}
When the junctions are not highly transparent, one can adopt a
perturbative approach by expanding in powers of the tunneling
hamiltonians $H_{Tj}$. For usual instances in which the time of
propagation between the tunnel junctions is larger than
$1/\Delta $, the contributions to the derivative in (\ref{cur})
are dominated by processes in which the Cooper pairs tunnel from
one of the superconductors to the 1D system, propagating then
to the other superconducting contact\cite{prox4}.
These first perturbative contributions give the result
\begin{eqnarray}
\lefteqn{ I (\chi )  \approx 8e t_1^2 t_2^2 \: k_B T
   \frac{d }{d \chi}
   \int_0^{1/k_B T} d\tau_1   \int_0^{\tau_1} d\tau_2
\int_0^{\tau_2} d\tau_3   \int_0^{\tau_3} d\tau_4 \langle
   \Psi^{\dagger}_{S1, \sigma}(d_1, -i\tau_1)
    \Psi^{\dagger}_{S1, -\sigma}(d_1,-i\tau_2) \rangle }
         \;\;\;\;\;\;\;\;\;\;\;\;\;\;
 \;\;\;\;\;\;\;\;\;\;\;\;\;\;\;\;\;\;\;\;\;\;\;\;\;\;\;\;  
 &  &   \nonumber              \\ 
   &   &    \langle  \Psi^{(a)}_{ \sigma}(d_1, -i\tau_1)
    \Psi^{(a)}_{-\sigma}(d_1, -i\tau_2)
   \Psi^{(a) \dagger}_{ \sigma'}(d_2, -i\tau_3)
\Psi^{(a) \dagger}_{ -\sigma'}(d_2, -i\tau_4) \rangle   \nonumber  \\  
  &   &  \langle \Psi_{S2, \sigma'}(d_2, -i\tau_3) 
\Psi_{S2, -\sigma'}(d_2, -i\tau_4) \rangle  
\label{pert}
\end{eqnarray}
where the statistical averages, at temperature $T$, are taken
over ordered products with respect to imaginary time $\tau $, in
the respective systems of the fields.

At low enough temperatures, we can approximate the statistical
averages at the boundary of the superconductors by delta
functions in imaginary time
\begin{equation}
\langle \Psi_{Sj, \sigma}(d_j, -i\tau_1)
              \Psi_{Sj, -\sigma}(d_j, -i\tau_2) \rangle \sim
 e^{i \chi_j}  N_j   \delta (\tau_1 - \tau_2 )
\end{equation}
$\chi_j $ being the phase of the order parameter.
Then the expression (\ref{pert}) is greatly simplified. The
maximum value of the supercurrent, that we call critical current
$I_c (T)$, is given by
\begin{equation}
 I_c (T)  \approx 2e N_1 N_2 t_1^2 t_2^2 
   \int_0^{1/k_B T} d\tau    
   \langle  \Psi^{(a)}_{ \sigma}(d_1, -i\tau )
      \Psi^{(a)}_{ -\sigma}(d_1, -i\tau )
  \Psi^{(a) \dagger }_{ \sigma'}(d_2, 0)
\Psi^{(a) \dagger }_{ -\sigma'}(d_2, 0)  \rangle  
\label{crit}
\end{equation}

We now make the passage to apply the above approach to the case
of the carbon nanotubes. The critical current in (\ref{crit}) is
given in terms of the propagator of the Cooper pairs along the
1D system. We need therefore to compute that object with the
fields corresponding to the linear branches crossing the Fermi
level of the carbon nanotubes, in the model governed by the
hamiltonian (\ref{ham2}).

The calculation of the 1D correlators can be accomplished by
making use of bosonization techniques\cite{emery,sol}.
For each linear branch of
the carbon nanotubes, one can define a boson operator
$\Phi^{(a)}_{r i \sigma}(x)$, related to the respective electron
density operator by
\begin{equation}
\partial_x \Phi^{(a)}_{r i \sigma}(x) = 
             2\pi \rho^{(a)}_{r i \sigma}(x)
\end{equation}
The bosonization method relies on the feasibility of expressing
the fermion fields in terms of the respective boson operators:
\begin{equation}
 \Psi^{(a)}_{\pm i \sigma }(x)   =   \frac{1}{\sqrt{\alpha_c }}
\exp \left( \pm i \Phi^{(a)}_{\pm i \sigma } (x) \right)
\label{bos}
\end{equation}
where $\alpha_c $ is a short-distance cutoff of the order of
$k_c^{-1}$ \cite{emery}.

In the case of the carbon nanotubes, the propagator that
enters in the computation of the critical current in (\ref{crit})
is of the type
\begin{equation}
 G (x,t)  \equiv \langle \Psi^{(a) \dagger}_{+ + \uparrow} (0,0)
\Psi^{(a) \dagger}_{- - \downarrow} (0,0) \Psi^{(a)}_{- -
\downarrow} (x,t) \Psi^{(a)}_{+ + \uparrow} (x,t)  \rangle
\label{prop}
\end{equation}
where the choice of linear branches accounts for the fact that
the Cooper pairs are formed with zero total momentum. The
evaluation of (\ref{prop}) can be done by using the bosonization
formulas (\ref{bos}) and passing to the combination of boson
fields that bring the hamiltonian (\ref{ham2}) into diagonal
form.

We observe then that, in the nanotube ropes,
the Coulomb interaction acts only on the combination of density
operators corresponding to the total charge density in the rope.
That is,
the repulsive interaction becomes sensible only in one out of
the $4n$ possible interaction channels, arising from the $n$
different metallic nanotubes and the degeneracy in the gapless
subbands and spin projections.

The $4n$ different channels can be classified into the sectors
corresponding to the sum and the difference of electronic charge
in the gapless subbands, and the rest which take into account
the spin densities for the two spin projections. The expression
of correlators like (\ref{prop}) factorizes into contributions
for each of the different sectors, in such a way that the
propagator for the Cooper pairs can be written in the form
\begin{equation}
   G (x,t)  =   \frac{1}{\alpha_c^2} C(x,t) \prod_{1}^{n} N(x,t)
\prod_{1}^{3n-1} F (x,t)
\label{fac}
\end{equation}
The first factor comes from the propagation of the total charge
density from all the metallic nanotubes, which has a purely
electrostatic interaction. The next $n$ equivalent factors stand
for the contribution of the charge mismatch in the gapless
subbands, which bears the effect of the attractive interaction
according to (\ref{ham2}).  The remaining $3n - 1$ factors
account for the free propagation of the rest of density
excitations, including the spin degrees of freedom and charge
degrees of freedom orthogonal to those affected by the
interaction.

Each of the factors in (\ref{fac}) corresponds to the
propagation of a boson field. At zero temperature, for instance,
all of them adhere to the common expression
\begin{equation}
X (x,t)  =   \exp \left( -\frac{1}{2n} \int_{0}^{k_c} dk
\frac{1}  {\mu (k)\: k} \left(1 - \cos (kx) \: \cos (\tilde{v}_F
kt) \right) \right)
\label{com}
\end{equation}
The respective quantities $\mu (k)$ arise in the change to the
variables that make the hamiltonian (\ref{ham2}) diagonal.  This
process also leads to a renormalization of the respective Fermi
velocities, given by $\tilde{v}_F (k) = v_F/\mu (k)$. In the
case of $C(x,t)$, we have $\mu (k) = 1/\sqrt{1 + 4n V_{C}(k)/\pi
v_F}$, while for $N(x,t)$ the quantity $\mu (k)$ does not depend
on the momentum, $\mu  = 1/\sqrt{1 - 4|g|/\pi v_F}$. For
the rest of excitations with free propagation, we simply have
$\mu (k) = 1$ and $\tilde{v}_F (k) = v_F$.

In practice, one is interested in the behavior of the
supercurrents at finite temperature, which requires moreover the
knowledge of the propagator (\ref{prop}) at imaginary time. For
this reason, it is more appropriate to deal with the
representation of objects like (\ref{com}) at finite temperature
in the Matsubara formalism, introducing the sum over discrete
frequencies $\omega_m = 2\pi m k_B T$ :
\begin{equation}
X (x,-i\tau)  =  \exp \left( -\frac{1}{2n} \int_{0}^{k_c} dk \;
\frac{2 k_B T}{v_F}      \sum_{m = -\infty}^{m = +\infty} \frac{1 -
\cos (kx) \: \cos (\omega_m \tau )} {(\omega_m /\tilde{v}_F)^2 +
k^2}  \right)
\label{temp}
\end{equation}

Thus, the framework developed is well-suited to discuss the
behavior of the supercurrents in carbon nanotubes depending on
the temperature $T$, the distance $d_1 - d_2$ between the
superconducting contacts, the number $n$ of metallic nanotubes
in a rope, and the competition between the repulsion given by the
Coulomb interaction and the effective attraction given by phonon
exchange.

\section{Comparison with experimental data}

In this section we establish the comparison with the
experimental measures of supercurrents in carbon nanotubes
reported in Ref. \cite{kas2}.
The supercurrents were observed in two
nanotube samples with different structure. One of them consisted
of a rope containing approximately 200 nanotubes, with a length
of about $1.7 \; \mu{\rm m}$. The other sample was made of seven
tubes merging at one of the ends into a single nanotube,
extending to a total length of $0.3 \; \mu{\rm m}$.  It has been
remarked in Ref. \cite{kas2}
that the critical current $I_c \approx 0.1
\; \mu{\rm A}$ measured in the thin rope was about 40 times
larger than expected within the conventional framework of the
proximity effect. The very flat shape of the critical
current as a function of the temperature was also surprising, as
its decay was only apparent in the neighborhood of the critical
temperature for the superconducting electrodes. The thick rope
showed a more conventional behavior, although the decay of the
supercurrent at the critical temperature of the electrodes was
not as sharp as expected\cite{kas2}.

We face the challenge of accounting for the experimental values
and features of the critical currents with our microscopic
model, that allows to describe ropes with any kind of nanotube
content. In this respect, our computational framework has only a
few adjustable parameters. The prefactors that affect the
strength of the critical current in (\ref{crit}) can be encoded
into the relative conductances $G_i$ at the tunnel junctions,
given (in units of $4e^2$) by $G_i = N_i t_i^2 /v_F$. The
critical current gets the right dimensions from the combination
$e v_F k_c$, that arises after setting $\alpha_c^{-1} = k_c$ and
performing the integral in (\ref{crit}) over the dimensionless
variable $\tilde{\tau} = v_F k_c \tau $. The expression of the
critical current becomes
\begin{equation}
   I_c (T)  \approx 2 G_1 G_2 e v_F k_c \int_0^{v_F k_c /k_B T}
d \tilde{\tau} C(L, -i\tilde{\tau}/v_F k_c) \prod_{1}^{n} N(L,
-i\tilde{\tau}/v_F k_c) \prod_{1}^{3n-1} F (L,
-i\tilde{\tau}/v_F k_c)
\label{crit2}
\end{equation}
where $L$ represents the distance between the superconducting
contacts.

Equation (\ref{crit2}) gives the critical current per metallic
nanotube in a rope. We observe that the order of magnitude is
given at short distances by the quantity $e v_F k_c$. It is
reasonable to set the short distance cutoff according to the
scale of the nanotube diameter, and we will take $k_c \approx
0.25 \; {\rm nm}^{-1}$ for typical nanotubes in a rope. This
sets a unit scale $e v_F k_c \approx 15 \; \mu{\rm A}$ for the
critical currents in carbon nanotubes.

\begin{figure}
\begin{center}
\mbox{\epsfxsize 8.5cm \epsfbox{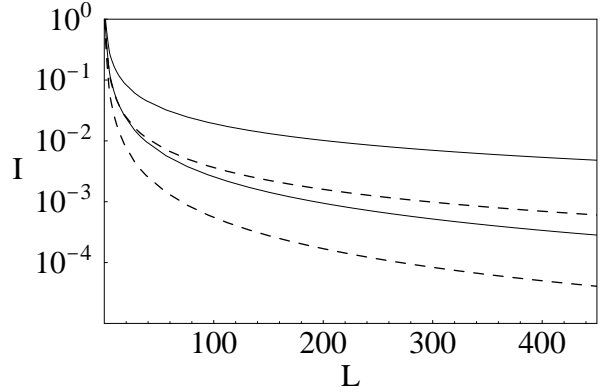}}
\end{center}
\caption{Plots of the critical current (measured in units of 
$e v_F k_c$) versus distance between the superconducting
contacts (in units of $k_c^{-1}$) at $T = 0$.
The full lines represent the results for ropes with a 
number of metallic nanotubes $n = 60$ (top) and $n = 2$ 
(bottom), taking $2 e^2/\pi^2 v_F = 1.0$ and a strength of
the effective attractive interaction $g/\pi v_F = 0.2$. 
The dashed lines represent the respective counterparts for the 
same values of $n$ in the absence of attractive interaction.}

\label{three}
\end{figure}

One has to bear in mind, however, that the critical currents
obtained from (\ref{crit2}) display a strong dependence on the
distance between superconducting contacts. Their decay for
increasing length $L$ can be very pronounced in ropes with a
small number of metallic nanotubes, where the electron
interactions are predominantly repulsive. The behavior at large
$L$ becomes much softer in the thick ropes, as shown in Fig.
\ref{three}. In general, the decay of the critical currents with
the length $L$ is not given by a simple power-law, since the
dependence of the Coulomb potential $V_C (k)$ on the momentum
prevents the perfect scaling of the observables at long
distances.

The results shown in Fig. \ref{three} allow us to draw the
comparison with the values of the critical currents at $T
\approx 0 \; {\rm K}$ reported in Ref. \cite{kas2}.
The appropriate
correspondence with the thin rope described there is established
by making the reasonable assumption that two out of the seven
nanotubes in the rope are metallic. Moreover, we determine the
theoretical value of the critical current by looking at the
decay along the distance between the contacts,
setting $L = 75/k_c $ for
the thin rope. For that choice of the parameters and taking a
strength $2e^2/\pi^2 v_F = 1.0$ for the Coulomb interaction, we
observe that the experimental value of the critical current
($I_c \approx 0.1 \; \mu{\rm A}$) can be matched by taking a
coupling for the attractive interaction $g/\pi v_F = 0.2$.
Actually, the value of the integral in (\ref{crit2}) at zero
temperature for $n = 2$ and $L = 75/k_c $ is $\approx 3.8 \times
10^{-3}$, which gives a critical current $ 2 I_c (0)
\approx 0.11 \; \mu{\rm A}$ (times $2 G_1 G_2$) after adding
the contributions of the two metallic nanotubes.

The precise determination of the critical current is affected by
the values of the relative conductances $G_1$ and $G_2$.  It is
however reassuring the fact that, once the interaction strengths
are fixed to the above mentioned values, one can reproduce
approximately the ratio between the critical currents observed
in the two different samples where the supercurrents have been
measured. In the case of the thick rope, we assume that it may
have a number of metallic nanotubes $n \approx 60$, and we make
the estimate of the critical current by computing the decay for
a $1.7 \; \mu{\rm m}$-long rope, which corresponds to $L =
425/k_c$ in our model. The value of the integral in
(\ref{crit2}) at zero temperature for $n = 60$ and $L = 425/k_c
$ is $\approx 5.1 \times 10^{-3}$. This leads us to assure that
the critical current in the thick rope should be about 40 times
greater than that in the thin rope, taking into account the
approximate ratio between the number of metallic nanotubes in
the two samples. This agrees well with the order of magnitude of
the experimental value $I_c \approx 2.5 \; \mu{\rm A}$ for the
thick rope reported in Ref. \cite{kas2}.

\begin{figure}
\begin{center}
\mbox{\epsfxsize 8.5cm \epsfbox{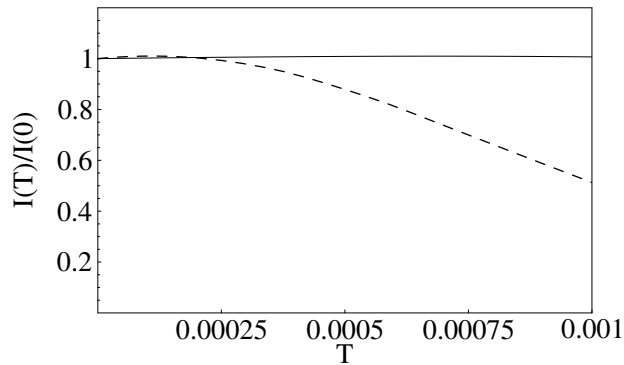}}
\end{center}
\caption{Plots of the normalized values of the critical current 
versus temperature (measured in units of $v_F k_c /k_B $) in a thin
rope with 2 metallic nanotubes, for $2 e^2/\pi^2 v_F = 1.0$, 
$g/\pi v_F = 0.2$, and respective nanotube lengths 
$L = 75/k_c$ (full line) and $L = 425/k_c$ (dashed line).}
\label{four}
\end{figure}

The other important check is that our model is able to account
for the dependence of the critical currents on temperature
observed experimentally. We have represented in Fig. \ref{four}
the behavior obtained from (\ref{crit2}), computing now the
critical current at a fixed distance $L = 75/k_c $, for the
thin rope with $n = 2$ (and keeping a suitable value of the
coupling $g/\pi v_F = 0.2$). The temperature is measured in
units of $v_F k_c /k_B \approx 1.2 \times 10^3 \; {\rm K}$.
Thus, in the case of the thin rope,
the transition temperature $T_c \approx 0.4 \; {\rm K}$
of the superconducting contacts corresponds to a value
$k_B T_c/v_F k_c \approx 3 \times 10^{-4}$ in the scale of Fig.
\ref{four}. We observe that the natural behavior of the critical
current is being flat over such range of small temperatures, in
agreement with the experimental measures in the thin rope
studied in Ref. \cite{kas2}.
The present results can be trusted while
remaining away from the critical temperature $T_c$ of the
superconducting electrodes, as we have not paid attention to the
dependence of their gaps on temperature. The sharp decay of the
critical current observed experimentally near $T_c$ is just the
natural consequence of approaching the superconducting
transition of the contacts.

We have to remark, however, that the flat behavior of the
critical current as a function of $T$ is a consequence of the
short distance along which it has been measured in the thin
rope. For long enough ropes, there is a
crossover temperature above which the behavior changes to a more
pronounced decrease of the critical current. This can be
appreciated in Fig. \ref{four}, where it has been represented
the critical current that would correspond to a thin rope with
$n = 2$ and length $L = 425/k_c $. In that case, the critical
current at zero temperature would be $2 I_c (0)
\approx 0.009 \; \mu{\rm A}$ (times $2 G_1 G_2$).
Most remarkably, an inflection
point is observed in the dependence on temperature.  On physical
grounds, such a crossover temperature marks the point above
which the thermal effects begin to affect significantly the
propagation of the Cooper pairs, inducing the pronounced decrease
of the supercurrent.

\begin{figure}
\begin{center}
\mbox{\epsfxsize 8.5cm \epsfbox{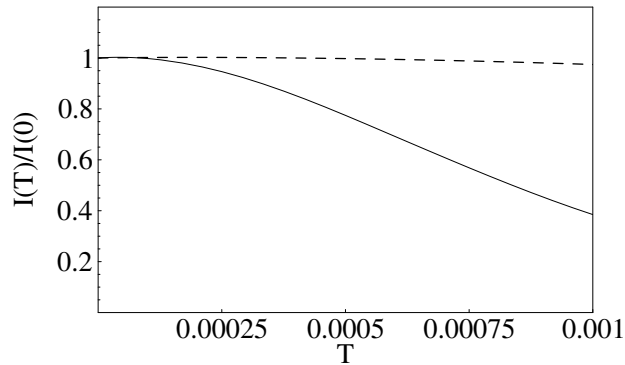}}
\end{center}
\caption{Plots of the normalized values of the critical current
versus temperature (measured in units of $v_F k_c /k_B$) in a thick
rope with 60 metallic nanotubes, for $2 e^2/\pi^2 v_F = 1.0$,
$g/\pi v_F = 0.125$, and respective nanotube lengths
$L = 425/k_c$ (full line) and $L = 75/k_c$ (dashed line).}
\label{five}
\end{figure}

The crossover temperature for the critical current is always
found below a scale which is of the order of $\sim v_F/k_B L$.
This is perfectly consistent with the behavior of the
$1.7 \; \mu{\rm m}$-long rope described in Ref. \cite{kas2},
where an inflection
point can be seen in the plot of the critical current as a
function of the temperature. We have represented in Fig.
\ref{five} the results for the critical current computed from
(\ref{crit2}), for $n = 60$ and $L = 425/k_c $.  Bearing in mind
that the temperature is measured in the figure in units of $v_F
k_c /k_B \approx 1.2 \times 10^3 \; {\rm K}$,
we observe that the inflection
point in the curve of $I_c (T)$ corresponds to a crossover
temperature $T \approx 0.7 \; {\rm K}$. This is in close
agreement with the experimental measures reported in Ref.
\cite{kas2}.

We have also represented in Fig. \ref{five} the behavior that
should be expected if the thick rope were shrinked down to a
length of $\approx 0.3 \; \mu{\rm m}$, keeping the same number
of metallic nanotubes. In that case, the critical current at
zero temperature could be as large as $n I_c (0) \approx 22.5
\; \mu{\rm A}$ (times $2 G_1 G_2$).
In agreement with our preceding discussion, the
crossover temperature would be shifted in the shorter sample to
$T \sim 5 \; {\rm K}$, a value that would be in general larger
than the transition temperature of the contacts and would make
then the crossover unobservable.

\section{Conclusions}

In this paper, we have developed a transport approach with
the aim of accounting for the properties of the supercurrents
measured in carbon nanotubes. Some important features of the
critical currents reported in Ref. \cite{kas2} cannot be explained
within the conventional picture of the proximity effect. In that
framework, the values of the supercurrent at zero temperature are
given by the expression $\pi \Delta/e R_N$, where $R_N$ is the
normal resistance of the junction. That estimate falls
short by a factor of 40, for instance, to reproduce the critical
current measured in the $0.3 \; \mu{\rm m}$-long rope considered
in Ref. \cite{kas2}. In the 1D approach we have adopted, the
values of the critical currents are determined instead by the
scale of the distance between the superconducting contacts,
provided that such a quantity is greater than the
coherence length in the superconductors.  The theoretical values
found in this way are in agreement with the experimental
measures for the two samples where supercurrents have been
reported in Ref. \cite{kas2}.

Even in a 1D framework, it is possible to make two different
descriptions depending on whether the contacts are supposed to
be perfectly transmitting or not. In the first case, the decay
of the critical current turns out to be independent of the
strength of the interactions at low temperatures\cite{prox3}.
Such an instance may not be the most appropriate to describe the
experimental conditions of the measurements of the proximity
effect described in Refs. \cite{kas2} and \cite{marc}. In
the latter, no evidence of supercurrents has been found for
individual nanotubes with a length of about $0.3 \; \mu{\rm m}$,
placed between Nb contacts.  One may think that the main
difference between the conditions of the experiments described in
Refs. \cite{kas2} and \cite{marc} lies in that, 
in the former, the samples were
made of ropes suspended between the electrodes. Such conditions
seem to be determinant for the observation of supercurrents,
what is supported by the fact that the normal 
resistances of the junctions were comparable in the two kind of
experiments. This should lead us to conclude that the
electron-electron interactions play an important role in the
observation of the proximity effect in the mentioned
experiments.
 
We have consequently adopted the point of view in which the
single-particle scattering dominates at the contacts between the
nanotubes and the electrodes. In these conditions, the values of
the critical current depend sensibly on the strength of the
Coulomb repulsion and the attractive interaction arising from
phonon exchange. In the thinner ropes, the repulsive interaction
does not suffer a significant reduction from the electrostatic
coupling among a small number of metallic nanotubes, and this
explains the comparatively small magnitudes of the supercurrents
in those instances. The value of the critical current $I_c
\approx 0.1 \; \mu{\rm A}$ for the very thin rope reported in
Ref. \cite{kas2} is actually dictated by the relatively small 
distance ($\approx 0.3 \; \mu{\rm m}$) between the contacts in 
that sample. We can foresee that, in a similar rope with only two
metallic nanotubes extending over a distance $L \approx 1.7 \;
\mu{\rm m}$ between superconducting contacts, the value of
the critical current should decay to a magnitude of $\sim
10^{-2} \; \mu{\rm A}$.

In the thicker ropes, larger supercurrents are expected as the
strength of the Coulomb repulsion is significantly reduced from
the interaction among the different metallic nanotubes. On the
other hand, it is known that the coupling resistance between
tubes in a rope is typically above $1 \; {\rm M}\Omega $, what
leads to assume that the conduction is confined to the
individual nanotubes.  Then, each metallic nanotube in the rope
provides a different channel for the propagation of Cooper
pairs, whenever there is good contact with the superconducting
electrodes. This condition can be difficult to ensure from an
experimental point of view, introducing some uncertainty about
the real number of metallic nanotubes that may contribute to
carry the supercurrent. We have seen, in particular, that it is
possible to account for the value of the critical current in the
thick rope described in Ref. \cite{kas2}, by adding the currents 
for the approximate number of metallic nanotubes in the rope.

Finally, let us remark that we have also given a satisfactory
description of the temperature dependence of the critical
currents reported in Ref. \cite{kas2}. The steady behavior of the
critical current in the thin rope studied there can be
understood as a consequence of the relatively short distance
between superconducting contacts for that sample. We have seen
that, for sufficiently long nanotubes, it is possible to observe
a crossover in temperature from the flat behavior to a
pronounced decay of the critical current, before the transition
temperature of the contacts is reached. The crossover has to take 
place at a temperature which is in general slightly below the 
scale of $v_F/k_B L$, and it manifest
by the presence of an inflection point in the plot of the
critical current as a function of the temperature. Such a
feature has been actually seen in the experimental
measures of the $1.7 \; \mu{\rm m}$-long rope studied in Ref.
\cite{kas2}. This comes as
a genuine signature of the propagation of the Cooper pairs
along the nanotubes, and stresses once again the suitability 
of the transport approach to account for the unconventional
properties observed in the carbon nanotubes.

\end{document}